\documentclass[epj]{webofc}
\usepackage[varg]{txfonts}   

\newcommand{\kcpppn}{K^{\pm} \to \pi^0\pi^0\pi^{\pm}~}
\newcommand{\kcpp}{K^{\pm} \to \pi^{\pm}\pi^0~}
\newcommand{\kcppg}{K^{\pm} \to \pi^{\pm}\pi^0\gamma}
\newcommand{\ke}{K_{e 2}}
\newcommand{\km}{K_{\mu 2}}

\newcommand{\RK} {\Gamma(K^{+}\to e^{+}\nu) /  \Gamma(K^{+}\to \mu^{+}\nu)}
\newcommand{\kcpnn}{K^{+} \to \pi^{+} \nu \bar{\nu}}

\woctitle{QCD@Work 2014}

\begin{document}
\title{NA62 experiment at CERN SPS}
%
%

\author{Venelin Kozhuharov\inst{1,2}\fnsep\thanks{\email{Venelin.Kozhuharov@cern.ch}\\ Speaker, for
 the NA62 Collaboration: 
 G. Aglieri Rinella, F. Ambrosino, B. Angelucci, A. Antonelli, G. Anzivino, R. Arcidiacono, 
I. Azhinenko, S. Balev, J. Bendotti, A. Biagioni
% ...
} 
}

\institute{Laboratori Nazionali di Frascati - INFN, 40 E. Fermi, 00044 Frascati (Rome), Italy 
\and
Faculty of Physics, University of Sofia ``St. Kl. Ohridski'', 5 J. Bourchier Blvd., 
1164 Sofia, Bulgaria 
          }

\abstract{%
   The NA62 experiment at SPS is a continuation of the long standing CERN kaon physics program. 
The high statistics and the unprecedent precision allow to probe the Standard Model and test 
the description of the strong interactions at low energy. The final results on the  the lepton 
universality test by measuring the ratio $R_K = \RK$ and the study of the 
$K^\pm\to\pi^{\pm}\gamma\gamma $ decay are presented.  The major goal of the NA62 
experiment is to perform a measurement of the  $Br(K^{+} \to \pi^{+} \nu \bar{\nu})$ with a 
precision of 10\% in two years of data taking. The detector setup together with the 
analysis technique is described. 
}
\maketitle
\section{Introduction}
\label{na62vvv:intro}

The high intensity approach of the fixed target experiments as opposed to the 
highest energy collisions provides a unique opportunity to address the 
Standard model through precision measurements. 

The phenomena in kaon physics allow to probe both the low energy behaviour 
of the strong interactions as well as the high energy weak scale through loop processes. 
Special attention should be given to the rare kaon decays since some of them 
could achieve sizeable contribution in the presence of New Physics.
In the Standard Model they are suppressed either by the necessity of
flavour changing neutral current transitions or due to helicity conservation.


\section{The NA62 experiment}
\label{na62vvv:na62}

The NA62 experiment is located at CERN North area and uses a primary proton beam 
from the SPS for the production of a secondary kaon beam.
Its first data taking took place in the 2007-2008 with the NA48/2 setup and was devoted 
to the study ot the $\ke$ decays.  
In 2009 the existing
experimental apparatus was dismantled in order to allow the construction 
of the new setup \cite{bib:na62tdr} devoted to the study of the $\kcpnn$ decay. 
In 2012 a technical run with beam was accomplished to study the performance 
of part of the NA62 subdetectors. At present the experiment is in its final 
construction and preparation phase and will have the pilot physics run in October 2014.

  \subsection{$\ke$ data taking setup}
 The kaon beam was formed by a primary 400 GeV/c proton beam extracted from SPS hitting a 
400 mm long beryllium target. 
The secondary particles were selected with a momentum of $(74 \pm 1.4)$ GeV/c with the possibility 
to use simultaneous or single positive and negative beams. The fraction of the kaons in the beam 
was about 6\% and they decayed in a 114 m long evacuated tank. 

The decay products were registered by the NA48 detector \cite{bib:na48}. The momentum 
of the charged particles was measured with resolution $\sigma(p)/p = (0.48 \oplus 0.009 p [GeV/c]) )\%  $ 
by a spectrometer consisting of four drift chambers separated by a dipole magnet. Precise time information and trigger
condition was provided by a scintillator hodoscope with time resolution of 150 ps which was followed by a quasi-homogeneous liquid
krypton electromagnetic calorimeter measuring photon and electron energy with 
resolution $\sigma(E)/E = 3.2\%/ \sqrt{E} \oplus 9\%/ E \oplus  0.42\%$ [GeV]. It was also able to 
provide particle identification based on the energy deposit by different particles with respect to 
their momentum.
 
A lead bar with 9.2 radiation lengths was placed in front of the LKr during 55\% of the data taking 
to study the muon misidentification probability. Data with three different beam conditions were collected - 65\% 
only with $K^+$ beam, 8\% only with $K^-$, and the rest with simultaneous beams

  \subsection{$\kcpnn$ experimental setup}
  The beam and the detector of the NA62 experiment for the $\kcpnn$  data taking 
 are dictated by the main goal - the study of the extremely rare decay $\kcpnn$. 
The protons intensity from the SPS will 
be increased by 30\% and the secondary positive beam will be with momentum $(75GeV/c  \pm 1\% )$. 
Its rate will be 
about 800 MHz and the decay volume is evacuated. 
The final beam line was tested during the technical run in 2012. 
\begin{figure}
 \includegraphics[width=\textwidth]{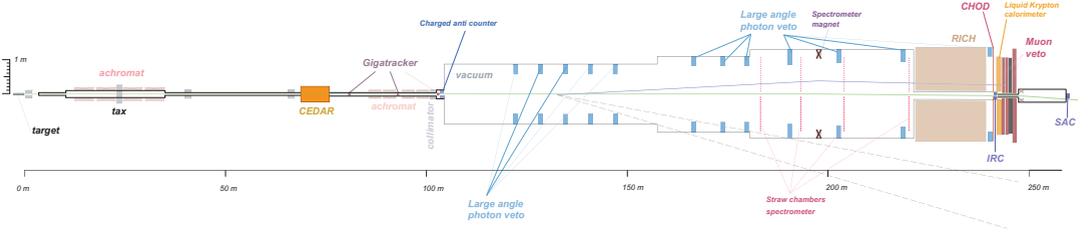}
 \caption{NA62 experimental layout}
 \label{fig:na62}
\end{figure}
  The major detector components are shown in fig. \ref{fig:na62} and are:

{\bf KTAG:} A hydrogen filled threshold Cherenkov counter used 
for positive kaon identification in the beam at a rate of 45 MHz.
Time resolution of 100 ps was achieved which is important for the suppression of accidental
background. 

{\bf Gigatracker:} Three stations of thin silicon pixel detectors for the measurement of kaon momentum,
 flight direction and time.
The expected resolutions on the measured quantities will be $\sigma(p_K)/p_K \sim 0.2\%$ on momentum, 
16 $\mu rad$ angular and  time resolution of 200 ps per station. 

{\bf Chanti:} Scintillating anticounters providing a veto against interactions of the beam particles. 

{\bf ANTI:} Twelve rings of lead glass counters surrounding the decay region and 
acting as photon veto detectors (LAV) \cite{bib:na62-lav}
for angles 
of the photons higher than 8.5 $mrad$ with respect to the kaon flight direction.

{\bf Straw spectrometer:} Four chambers of straw tubes separated by the 
MNP33 dipole magnet will be operated in vacuum in order to provide
momentum resolution of $\sigma(p)/p = (0.3 \oplus 0.008 p[GeV/c]) )\%  $ with a minimal material budget. 

{\bf RICH:} A ring imaging Cherenkov detector will measure the 
velocity of the charged particles allowing to separate 
pions from muons and will provide time resolution better than 100 ps \cite{bib:rich}. 

{\bf CHOD:} A plastic scintillator charged hodoscope will be used in the trigger. 

{\bf IRC and SAC:} Shashlyk type veto detectors covering photon angles down to zero (SAC) and also serving 
to veto converted in the upstream material photons (IRC). 

{\bf LKr:} The NA48 liquid krypton calorimeter with renewed readout electronics will serve as a 
photon veto for photons with angles from 1.5 to 8.5 $mrad$ with inefficiency less than 
$10^{-5}$ for photon with energies above 10 GeV. 

{\bf MUV:} Three muon veto stations based on iron and scintillator sandwich 
will provide separation between pions and muons better than $10^{-5}$. 

Both KTAG and Gigatracker are exposed to the full 800 MHz hadron 
beam while the rate seen by the downstream detectors is at most 10 MHz.

\section{Probing the lepton universality with $K^{\pm} \to l^{\pm}\nu$ decays}

Within the Standard Model the dilepton charged pseudoscalar meson decays proceed as tree level 
processes through a W exchange. 
However, the helicity conservation leads to a strong suppression of the electron mode.
The Standard Model (SM) expression for the ratio $R_K= \Gamma(Ke2) / \Gamma(K\mu 2)$ is a function 
of the masses of the participating particles and is given by
\begin{equation}
R_K=\frac{m_e^2}{m_{\mu}^2} \left(  \frac{m_K^2 - m_e^2}{m_K^2 - m_{\mu}^2} \right)   (1+\delta R_K),
\end{equation}
where the term $ \delta R_K = -(3.79 \pm 0.04) \% $ represents the radiative corrections. 
In the ratio $R_K$ the theoretical uncertainties on the hadronic matrix element cancel 
resulting in an extremely precise prediction $R_K = (2.477 \pm 0.001) \times 10^{-5} $ \cite{ke2-thnew}. 

Due to the impossibility to distinguish the neutrino flavour the experimentally measured ratio is 
sensitive to possible lepton flavour violation effects. In particular, 
various LFV extensions of the SM (MSSM, different two Higgs doublet models) 
predict constructive or destructive contribution to $R_K$ as high as 1 \% \cite{Masiero}.

Experimentally, the ratio $R_K$ can be expressed as 
\begin{equation}
R_K =
\frac{1}{D}\cdot
\frac{N(\ke)-N_B(\ke)}
{N(\km) - N_B(\km)}
\cdot
\frac{A(\km)\times\epsilon_{\mathrm{trig}}(\km)\times f_\mu}
{A(\ke)\times\epsilon_{\mathrm{trig}}(\ke)\times f_e} \cdot \frac{1}{f_{LKr}},
\label{RKexp}
\end{equation}
where  $N(K_{\ell 2})$, $\ell=e,\mu$ is the number of the selected $\ke$ 
and $\km$ candidates,  
$N_B(K_{\ell 2})$ is the number of expected background events, 
$f_\ell$ is the efficiency for particle identification, 
$A(K_{\ell 2})$ is the geometrical efficiency for registration obtained from Monte Carlo simulation, 
$\epsilon_{\mathrm{trig}}$ is the trigger efficiency, $D=150$ is the downscaling factor for $\km$ events and $f_{lkr}$ is the global efficiency of the LKr readout. Both $f_\ell$ and $\epsilon_{\mathrm{trig}}$ are higher than 99\%. 

The analysis was performed in individual momentum bins for all the four different data samples - $K^+(noPb)$, $K^+(Pb)$, $K^-(noPb)$, $K^-(Pb)$ - resulting into 40 independent values for the $R_K$. 

The similarity between the two decays allowed to exploit systematics cancellations in the ratio by using common selection criteria. 
The events were required to have only one reconstructed charged track within the detector geometrical 
acceptance with momentum in the interval $13 ~GeV/c < p < 65 ~GeV/c$ and consistent with kaon decay. 
The background was additionally suppressed by vetoing events with clusters in the LKr not associated with the track with energy more than 2 $GeV$. 
The particle identification was based on the $E/p$ variable, where $E$ is the energy deposited in the LKr and $p$ is the momentum measured by the spectrometer. 
It had to be close to one for electrons and less than 0.85 for muons. 
Under the assumption of the particle type the missing mass squared was calculated $M_{miss}^2 = (P_K - P_l)^2$, 
where $P_K$ ($P_l$) is the kaon (lepton) four momentum. A momentum dependent cut on the $M_{miss}^2$  was used.

The dominant background contribution in the $\ke$ sample was identified 
to come from $\km$ events with muons leaving all their energy in the electromagnetic calorimeter. 
The two decays are well separated below 35 $GeV/c$ track momentum but completely overlap kinematically for higher values. 
In order to select clean muon samples the data with the Pb wall was used. 
The study of the probability of a muon faking an electron 
was performed separately in the different momentum bins. 
The estimated background from $\km$ events was $(5.64 \pm 0.20)\%$ 
with uncertainty dominated by the statistics used for the determination of $P_{\mu e}^{Pb}$.

At low track momentum the most significant background source was identified to be the muon halo. 
The total background after taking into account also the structure dependent and 
the interference part of the $K^{\pm} \to e^{\pm} \nu \gamma $ decays 
 was found to be $(10.95 \pm 0.27)\%$. 

\begin{figure}[!htb]
    \resizebox{0.42\textwidth}{!}{\includegraphics[width=0.42\textwidth]{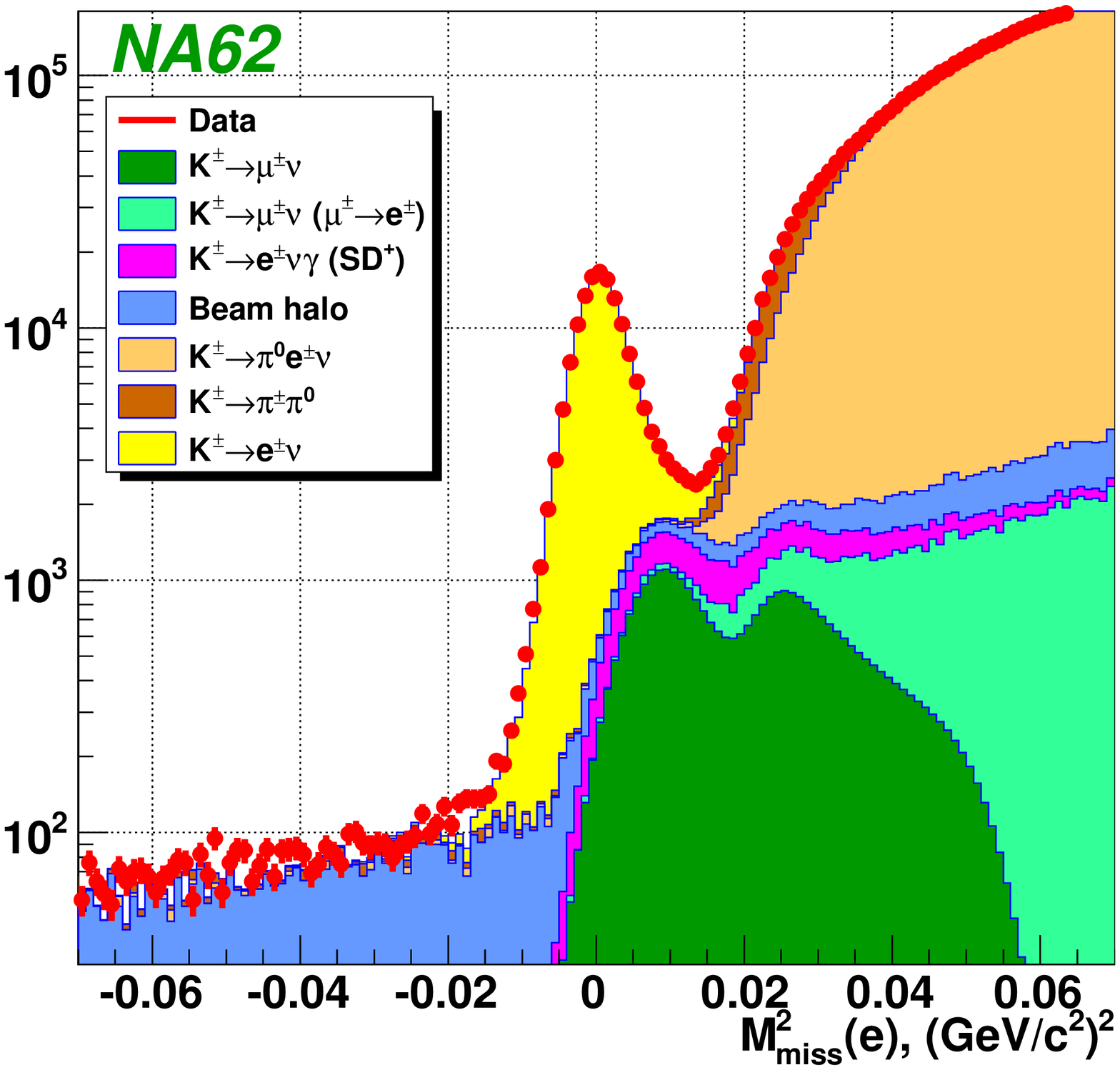}}
    \resizebox{0.49\textwidth}{!}{\includegraphics[width=0.49\textwidth]{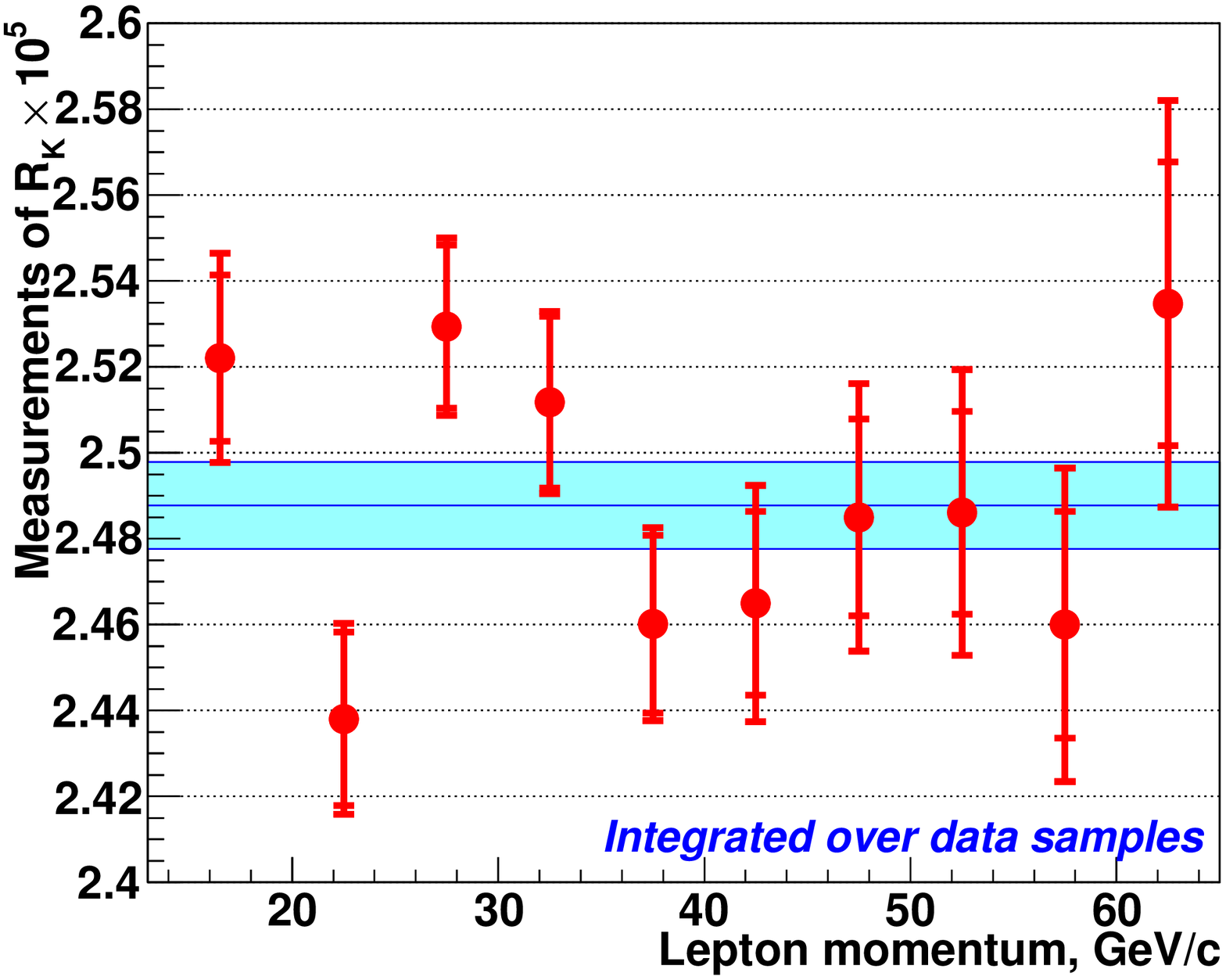}}
\put(-300,165){ \bf (a)}
\put(-100,165){ \bf (b)}
    \caption{ {(a) Squared missing mass distribution for the reconstructed data events (dots) together with the major background contribution. (b) $R_K$ in bins of track momentum integrated over data samples. The horizontal dashes represent the statistical only and the total error.} \label{fig:ke2} }
\end{figure}

The missing mass distribution for the reconstructed $\ke$ data events together with the simulation of the 
signal and backgrounds are shown in fig. \ref{fig:ke2} (a). A total of $145958$ $\ke$ and $4.28\times 10^{7}$  $\km$ 
candidates were reconstructed. 
The value of $R_K$ in the in individual momentum bins integrated over the data samples is 
shown in fig. \ref{fig:ke2} (b). 
The final result was obtained by a fit of the 40 independent $R_K$ values and is \cite{bib:na62-ke2}
\begin{equation}
  R_K = (2.488 \pm 0.007_{stat} \pm 0.007_{syst})\times 10^{-5}.
\end{equation}
It is consistent with the Standard Model prediction and with the present PDG value \cite{bib:ke2-pdg} 
and supersedes the previous \cite{bib:na62-ke2prel}  NA62 result.

\section{Tests of the ChPT with $K^{\pm}\to\pi^{\pm}\gamma\gamma$}

  Within the ChPT the lowest order terms contributing to the the 
decay $K^{\pm}\to \pi^{\pm}\gamma\gamma$ are of order $O(p^4)$ \cite{bib:pigg-th1}. 
They represent the pion and the kaon loop amplitudes depending on a 
single unknown constant $\hat{c}$ and a pole amplitude contributing of the 
order of 5\% to the final decay rate. 
Higher order corrections ($O(p^6)$) have shown to change the 
decay spectrum significantly leading to a non vanishing differential decay rate at zero 
diphoton invariant mass \cite{bib:pigg-th2}. 
The predicted branching fraction for the $K^{\pm}\to \pi^{\pm}\gamma\gamma$ decay 
is $\sim 10^{-6}$.
It has been studied by the BNL E787 experiment 
which observed 31 decay candidates \cite{bib:pigg-bnl} and by 
the NA48/2 experiment, which identified 149 event candicates in a three day 
run with minimum bias trigger in 2004 \cite{bib:na48-pigg}.

\begin{figure}[!htb]
    \resizebox{0.49\textwidth}{!}{\includegraphics[width=0.49\textwidth]{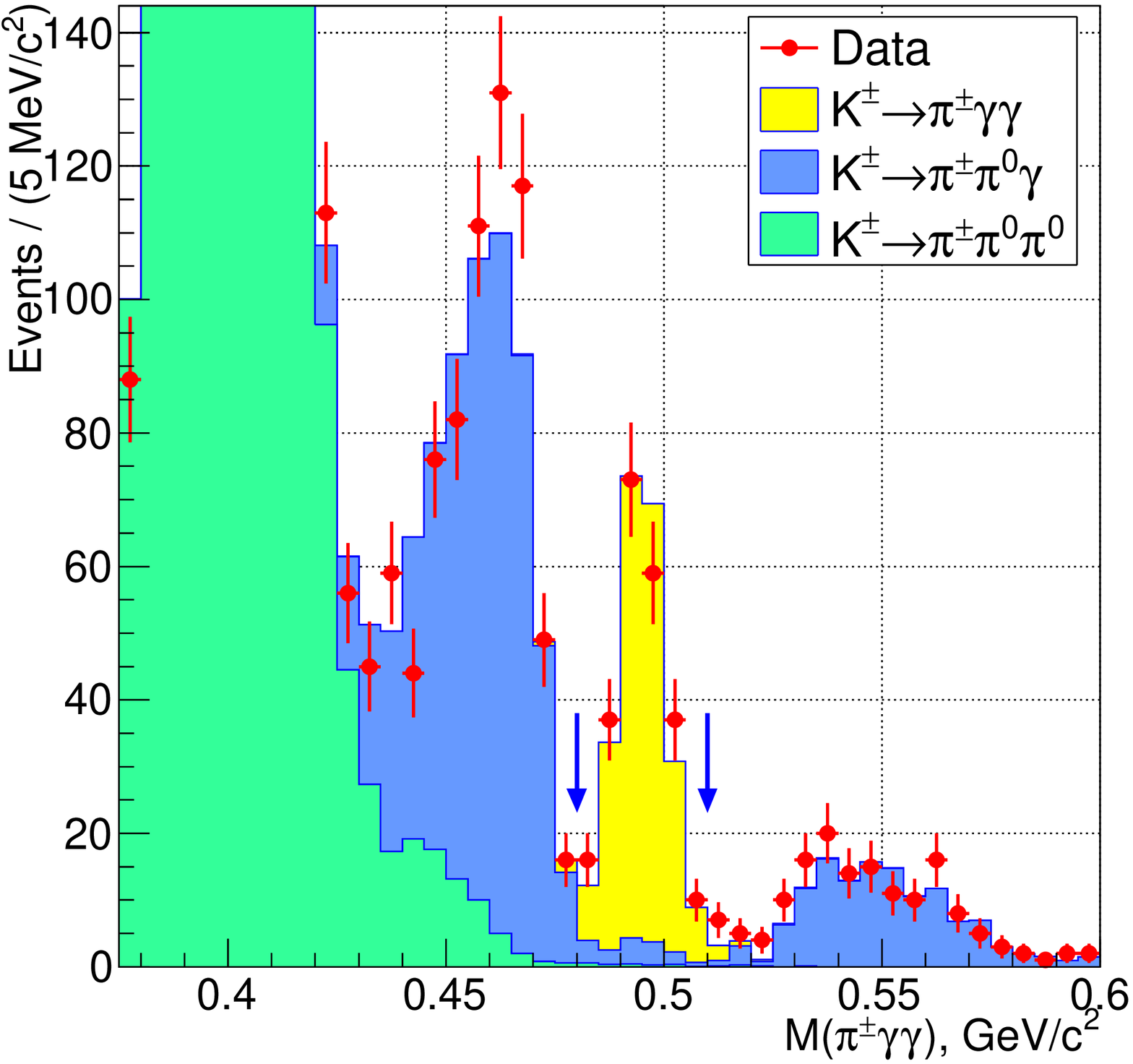}}
    \resizebox{0.49\textwidth}{!}{\includegraphics[width=0.49\textwidth]{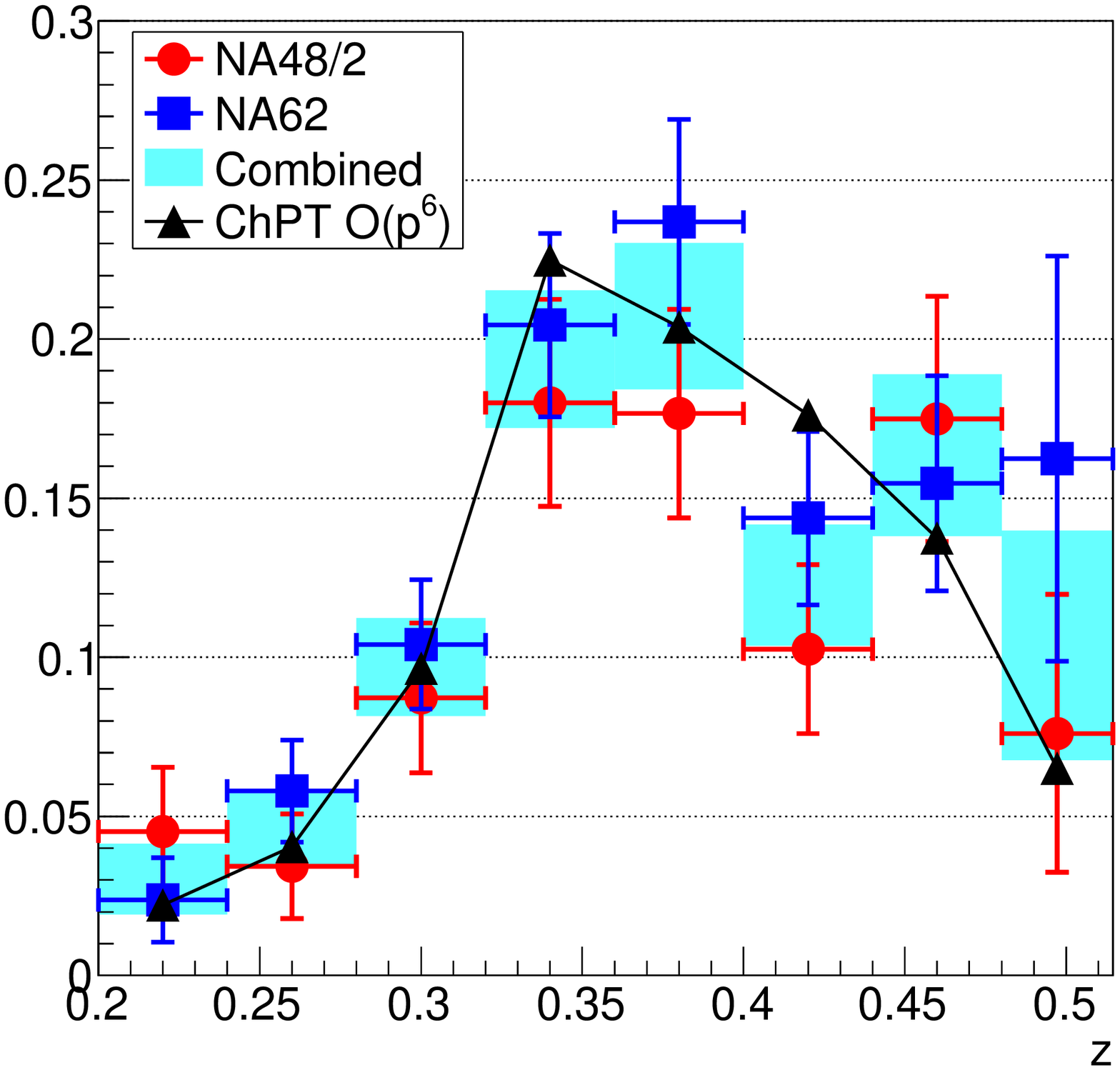}}
\put(-300,190){ \bf (a)}
\put(-100,190){ \bf (b)}
    \caption{ {(a) Invariant mass for the reconstructed data events together with the expected background contribution. (b) The data spectra of the z kinematic variable of the NA48/2 and NA62 data samples together with their average. The theoretical points for the obtained value of $\hat{c}$ from the fit to the data are also shown.
}  \label{fig:kpgg} }
\end{figure}

The data used for the study $K^{\pm}\to \pi^{\pm}\gamma\gamma$ was collected during the 2007 
run with a trigger with an effective downscaling of about 20.
The signal events were required to have $ z= (m_{\gamma\gamma} / m_K)^2 > 0.2$ in order to suppress the background from
$\kcpp$ decays. 
The reconstructed kaon invariant mass for the event candidates is shown in fig. \ref{fig:kpgg} (a).

A total of 232 event candidates were selected, with backgrounds
contaminations of 7\% from $\kcpppn$ and $\kcppg$ with merged clusters. 
The $z$ spectrum for the NA62 event candidates is shown in fig. \ref{fig:kpgg} (b)
together with the NA48/2 data and the avaraged values. The acceptance and the 
background contributions were also calculated separately in each $z$ interval.
This allowed to extract a combined NA62 + NA48/2 model independent branching 
ratio  \cite{bib:na62-pigg}

\begin{equation}
 Br(K^{\pm}\to \pi^{\pm}\gamma\gamma )_{MI,z>0.2} = (0.965 \pm 0.061_{stat} \pm 0.014_{syst} )\times 10^{-6}.
\end{equation}

The extraction of the $\hat{c}$ was based on a likelihood fit to the data
for both $O(p^4)$ and $O(p^6)$ parametrizations. 
The final results for $\hat{c}$ both for NA62 data alone and combined with NA48/2 are shown in table \ref{tab:pigg-prel}.

\begin{table}[!htb]
  \caption{Results for $\hat{c}$ within $O(p^4)$ and $O(p^6)$ ChPT parametrization. }
  \label{tab:pigg-prel}
  \begin{tabular}{c|c|c}
    \hline
  ChPT parametrization         &      NA62 results            &    Combined NA48/2 + NA62 results              \\
    \hline
 $O(p^4)$     &  $\hat{c} = 1.93 \pm 0.26_{stat} \pm 0.08_{syst}$  & $\hat{c} = 1.72 \pm 0.20_{stat} \pm 0.06_{syst}$   \\
 $O(p^6)$     &  $\hat{c} = 2.10 \pm 0.28_{stat} \pm 0.18_{syst}$  & $\hat{c} = 1.86 \pm 0.23_{stat} \pm 0.11_{syst}$ \\
    \hline
  \end{tabular}
\end{table}

The obtained result doesn't allow to discriminate between the $O(p^4)$ and $O(p^6)$  parametrization. 
Using the $O(p^6)$ description and the obtained combined NA48/2 + NA62 value for $\hat{c}$ 
the branching ratio in the  full $z$ kinematic region was found to be 

\begin{equation}
 Br(K^{\pm}\to \pi^{\pm}\gamma\gamma )_{O(p^6)} = (1.003 \pm 0.051_{stat} \pm 0.024_{syst} )\times 10^{-6}.
\end{equation}

\section{Measurement of $BR(\kcpnn)$ with 10\% precision}

Among the rare kaon decays the transitions $K \rightarrow \pi \nu \bar{\nu}$ are extremely attractive. 
They proceed as FCNC and their branching fractions are theoretically very clean since the hadronic 
matrix element can be obtained by the isospin symmetry of the strong interactions from the leading 
decay $K^+ \rightarrow \pi^0 e^+ \nu$ \cite{ISOSPIN_RELATION}. For the charged kaon mode the NNLO 
calculations give $Br(\kcpnn) = (7.81\pm0.80)*10^{-11}$ \cite{pnn-th}. 

Presently seven $\kcpnn$ events have been observed by the E787 and E949 collaborations 
in a stopped kaon experiment \cite{BNL_BR} leading to $Br(\kcpnn) = (1.73_{-1.05}^{+1.15})\times 10^{-10}$. 
This value is twice the SM prediction but still compatible with it due to high uncertainty. 
The decay $\kcpnn$ is very sensitive to New Physics models where the theoretical predictions vary over an order of magnitude. 
Thus measuring $Br(K^+ \rightarrow \pi^+ \nu \bar{\nu})$ could also help to distinguish between the different 
types of new physics \cite{PINN_BSM} once it is discovered.

\begin{figure}[!htb]
\begin{center}
\includegraphics[width=0.8\textwidth]{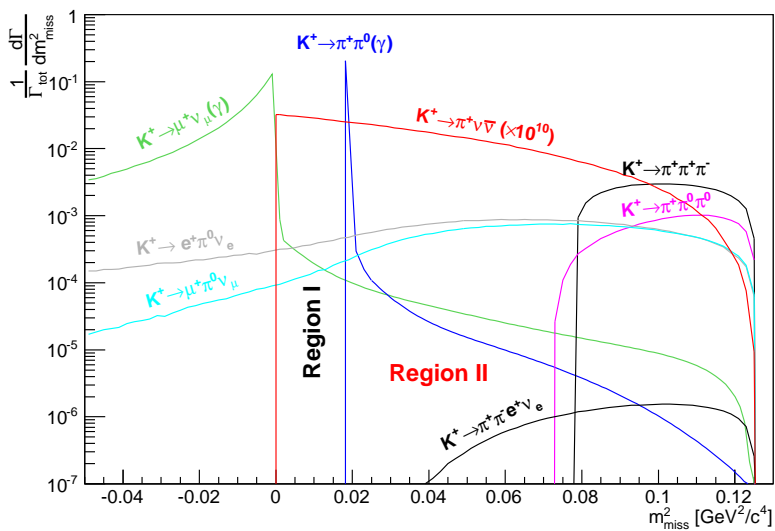}
\end{center}
    \caption{ { Squared missing mass distribution for the $\kcpnn$ and the main
charged kaon decay modes. The signal is multiplied by a factor $10^{10}$ while the backgrounds 
are according to their branching ratios. 
}     \label{fig:kna62} }
\end{figure}

The measurement which NA62 collaboration aims to perform is based on the kaon decay in flight. 
In order to identify the signal events and suppress the background the three techniques - 
kinematics, particle identification and vetoing will be exploited.
Since  there is only one observable particle in the final state the kinematics
variable considered for the separation of the decay is the missing mass squared under 
pion hypothesis for the charged track. It is defined as 
\begin{equation}
 m_{miss}^2 \simeq m_K^2\left(1-\frac{|P_{\pi}^2|}{|P_{K}^2|}\right)  + m_{\pi}^2\left(1-\frac{|P_{K}^2|}{|P_{\pi}^2|}\right) - |P_{K}||P_{\pi}|\theta_{\pi K}^2 
\end{equation}
With the planned Gigatracker and Straw spectrometer the expected resolution on the missing mass squared is $0.001 ~GeV^2/c^4$. 
The signal region is defined by the edges of the kinematically limited in the $m_{miss}^2$ 
distribution kaon decays: $K^+ \to \mu^+ \nu$, $K^+ \to \pi^+ \pi^0$ and $K^+\to \pi^+\pi^+\pi^-$, as shown in fig. \ref{fig:kna62}(b). 

The detector is hermetic for photons with angles up to 50 $mrad$ 
originating from $\pi^0$ in the decay region and the
overall photon veto system composed from ANTI, LKR, SAC, and IRC provides an inefficiency 
less than $10^{-8}$ for a $\pi^0$ coming from $K^+\rightarrow \pi^+\pi^0$ decay.
 
Decays with muons in the final state (like $K^+\rightarrow \mu^+ \nu$, $K^+\rightarrow \pi^+\pi^-\mu^+ \nu$) 
will be suppressed using the muon-pion identification based on RICH and MUV for which the total 
inefficiency should be less than  $5 \times 10^{-6}$.

The presented setup and analysis strategy will allow NA62 experiment to collect O(100) 
events in two years of data taking. 
The construction is advanced for the start of the experiment in October 2014.

\section{Rare and new physics processes with NA62}

NA62 experiment will accumulate the highest available charged kaons statistics. 
Combined with the excellent veto efficiency, particle identification capabilities, and 
ultimate momentum and energy resolution this turns the NA62 experiment into a multipurpose
facility able to execute a diverse physics program devoted to rare processes. Among them are the 
lepton flavour violation kaon decays, searches for new particles 
(including dark photons and heavy neutrinos), searches for 
forbidden kaon and pion decays. 
In addition, an extensive and high precision study of the $K^+ \to \pi^0\pi^0 l^+ \nu$ and 
$K^{\pm}\to\pi^{\pm}\gamma\gamma$ decays is also under consideration.

\section{Conclusions}

The rare kaon decays continue to provide a valuable input to the high energy physics. 
The NA62 experiment with its huge statistics, excellent resolution, particle identification, and 
hermeticity is the future laboratory for charged kaon physics. 
Currently a four per mile measurement of $R_K$ and a new higher statistics study
of the $K^{\pm}\to \pi^{\pm}\gamma\gamma$ have been performed. 
The next step is the study of $\kcpnn$ decay and the measurement of the CKM matrix element
$V_{td}$ with a 10\% precision. 
The general purpose experimental setup could allow to study  
other rare decays, including lepton flavour and lepton number violating ones.


\begin{thebibliography}{99}

\bibitem{bib:na62tdr} 
F. Hahn {\it et al.} [NA62 Collaboration], 
http://cds.cern.ch/record/1404985.

\bibitem{bib:na48} {V. Fanti {\it et al.} [NA48 Collaboration],}
    {Nucl. Instrum. Meth. A} {\bf 574} (2007) {433}.

\bibitem{bib:na62-lav}
  P.~Massarotti {\it et al.},
  PoS ICHEP {\bf 2012}, 504 (2013).

\bibitem{bib:rich} 
{B. Angelucci {\it et al.}}
  {Nucl.\ Instrum.\ Meth.\ A} {\bf 621} , {205} (2010)



\bibitem{ke2-thnew} {V. Cirigliano, I. Rosell,}
  {Phys. Rev. Lett.} {\bf 99}, 231801 (2007).


\bibitem{Masiero} {A. Masiero {\it et al.}, }
{Phys. Rev. D} {\bf 74}, 011701 (2006).

\bibitem{bib:na62-ke2} C. Lazzeroni {\it et al.} [NA62 Collaboration],  Phys. Lett. B {\bf 719}, 326 (2013). 

\bibitem{bib:na62-ke2prel} {C.~Lazzeroni {\it et al.}  [NA62 Collaboration],}
{Phys.\ Lett.\ B} {\bf 698}, {105} (2011).


\bibitem{bib:ke2-pdg} {J. Beringer  {\it et al.} [Particle Data Group],}
{Phys. Rev. D} {\bf 86} {2012} {010001}.



\bibitem{bib:pigg-th1} 
{G. Ecker, A. Pich and E. de Rafael}
{Nucl. Phys. B} {\bf 303}, {665} (1988).


\bibitem{bib:pigg-th2} 
{G. D’Ambrosio and J. Portoles}
{Phys. Lett. B} {\bf 386}, {403}  (1996).

\bibitem{bib:pigg-bnl} 
{P. Kitching {\it et al.} [E787 Collaboration]}
{Phys. Rev. Lett.} {\bf 79}, 4079 (1997).

\bibitem{bib:na48-pigg}
J.~R.~Batley {\it et al.}  [NA48/2 Collaboration], 
  Phys.\ Lett.\ B {\bf 730}, 141 (2014). 

\bibitem{bib:na62-pigg}
  C.~Lazzeroni {\it et al.}  [NA62 Collaboration],
  Phys.\ Lett.\ B {\bf 732}, 65 (2014).

\bibitem{ISOSPIN_RELATION}
 {F. Mescia and C. Smith}
{Phys. Rev. D} {76}, 034017 (2007).
\bibitem{pnn-th}
{J. Brod, M. Gorbahn and E. Stamou}
{Phys. Rev. D} {} 034030 (2011).

\bibitem{BNL_BR}
{A. Artamonov  {\it et al.} [E949 Collaboration]}
{Phys. Rev. D} {\bf 79}, 092004 (2009).


\bibitem{PINN_BSM}
{D. Straub}, 
arXiv:1012.3893 [hep-ph].


\end{thebibliography}
\end{document}